\documentclass[12pt, preprint]{aastex}
\usepackage{psfig,lscape}
\received{2003 February 9}
\begin{document}

\def\ft#1{\tablenotemark{\;\rm #1}}
\def\cft#1{$^{\;\rm #1}$}

\title{EMISSION-LINE DIAGNOSTICS OF THE CENTRAL ENGINES OF WEAK-LINE
RADIO GALAXIES}

\shorttitle{Emission-line Diagnostics of WLRGs}

\author{Karen T. Lewis\altaffilmark{1}, Michael
Eracleous\altaffilmark{1,2}, \& Rita M. Sambruna\altaffilmark{3}}

\altaffiltext{1}{Department of Astronomy and Astrophysics, The
Pennsylvania State University, 525 Davey Laboratory, University Park,
PA 16802, e-mail: {\tt lewis, mce@astro.psu.edu}}

\altaffiltext{2}{Visiting astronomer, Kitt Peak National Observatory
and Cerro Tololo Interamerican Observatory, which are operated by
AURA, Inc.  under a cooperative agreement with the National Science
Foundation.}

\altaffiltext{3}{Department of Physics and Astronomy and School of 
Computational Sciences, George Mason University, 4400 University Dr.
Fairfax, VA 22030, e-mail: {\tt rms@physics.gmu.edu}}

\begin{abstract}

A handful of well-studied Weak-Line Radio Galaxies (WLRGs) have been
traditionally classified as Low Ionization Nuclear Emission-line
Regions (LINERs), suggesting that these two groups of AGNs might be
related. In this paper, we present new optical emission-line
measurements for twenty Weak-Line Radio Galaxies which we supplement
with measurements for an additional four from the
literature. Classifying these objects by their emission-line ratios,
we find that 50\% of the objects are robustly classified as LINERs
while an additional 25\% are likely to be LINERs.  Photoionization
calculations show that the Spectral Energy Distribution of the
well-studied WLRG 3C 270 (NGC 4261) is able to produce the observed
emission-line ratios, but only if the UV emission seen by the narrow
emission-line gas is significantly higher than that observed, implying
A$_{\rm V}$ = 2.5 -- 4.2 magnitudes along our line of sight to the
nucleus. From the photoionization calculations, we find that the
emission-line gas must have an ionization parameter between
10$^{-3.5}$ and 10$^{-4.0}$ and a wide range in hydrogen density
(10$^{2}$ -- 10$^{6}$ cm$^{-3}$) to reproduce the measured
emission-line ratios, similar to the properties inferred for the
emission-line gas in LINERs. Thus, we find that properties of the
emission-line gas as well as the underlying excitation mechanism are
indeed similar in LINERs and WLRGs. By extension, the central engines
of accretion-powered LINERs and WLRGs, which do host an accreting
black hole, may be qualitatively similar.

\end{abstract}


\section{INTRODUCTION}

Early optical studies of powerful, extragalactic radio sources
revealed that the galaxies associated with them often showed strong
emission lines in their optical spectra \citep{s65}. To better
understand the correlations between radio and optical line emission,
\citet{t93} collected optical spectra of a complete sample of southern
2 Jy radio galaxies and quasars, derived from the \citet{wp85} 2.7-GHz
sample. Analyzing these data, \citet{t98} recovered the previously
known correlation between luminosity of the [\ion{O}{3}]$\lambda 5007$
emission line (L$_{\rm [O~III]}$) and the total radio luminosity
(L$_{\rm rad}$) \citep[see, for example,][]{RS91}.

However, \citet{t98} also found a large population of radio galaxies,
dubbed Weak-Line Radio Galaxies (WLRGs), that had very weak
[\ion{O}{3}] emission (i.e. equivalent width of [\ion{O}{3}] less than
10~\AA), despite being powerful radio sources. In fact, all of the FR
I \citep{fr74} sources in the sample were classified as WLRGs. This
was not surprising, since \citet{hl79} noticed that strong narrow
optical emission-lines were almost always associated with FR II
galaxies. Except for one object, Hydra A, \citet{t98} obtained only
upper limits to L$_{\rm [O~III]}$ for the FR I radio galaxies. Thus,
it was difficult to determine whether the FR I WLRGs followed the same
L$_{\rm [O~III]}$ - L$_{\rm rad}$ correlation as the higher luminosity
radio galaxies, or whether they truly had anomalously low [\ion{O}{3}]
luminosities. Hydra A did fall well below the correlation and the
measured ratios of [\ion{O}{3}]/[\ion{O}{2}] and [\ion{O}{3}]/H$\beta$
(both $<$ 1) indicated that the emission-line gas in this object might
be in a low-ionization state.

Nearly half of the WLRGs, however, were FR II radio galaxies. These FR
II sources may be related to the ``low-excitation'' FR II radio
galaxies noticed by both \citet{L94} and \citet{hl79}, in which
high-ionization lines, such as [\ion{O}{3}], were very weak compared
to the hydrogen lines.  When \citet{t98} restricted their sample to
the redshift range 0.1 --0.7, in which all the objects had measured
luminosities of [\ion{O}{3}], [\ion{O}{2}], and H$\beta$, the WLRGs
clearly stood out; L$_{\rm [O~III]}$ was an order of magnitude lower
than that of other radio galaxies with a similar L$_{\rm rad}$. On the
other hand, L$_{\rm [O~II]}$ and L$_{\rm H\beta}$ in these WLRGs were
not as drastically reduced, lending further support to the idea that
these FR II sources are ``low-excitation'' sources, as was suggested
by earlier studies. 

Independently of the above study, in a large X-ray survey of
radio-loud Active Galactic Nuclei (AGNs), \citet{s99} found a similar
correlation between the 2--10 keV X-ray luminosity (L$_{2-10 {\rm
keV}}$) and the 5 GHz radio lobe power (L$_{5 {\rm GHz}}$). Again, the
WLRGs in the sample stood out.  Compared to objects with similar L$_{5
{\rm GHz}}$, the WLRGs had L$_{2-10 {\rm keV}}$ as much as two orders
of magnitude lower than expected from the correlation. Furthermore,
the X-ray spectra of WLRGs were harder than those of other objects,
suggesting that WLRGs might be a distinct group of AGN.

The relative weakness of the high-ionization [\ion{O}{3}] emission
line in WLRGs suggests that these objects might be related to Low
Ionization Nuclear Emission Regions (LINERs; Heckman 1980, hereafter
H80). H80 defined this class of objects with two line intensity
ratios: [\ion{O}{2}]$\lambda 3727$/[\ion{O}{3}]$\lambda 5007
\ge$ 1 and [\ion{O}{1}]$\lambda 6300$/[\ion{O}{3}]$\lambda 5007 \ge 1/3$ (the
H80 criterion). Indeed, several objects traditionally classified as
LINERs (e.g. 3C 270, NGC 6251, and Hydra A) were classified by
\citet{t98} as WLRGs.

The optical emission-line properties of LINERs can be reproduced by a
variety of mechanisms, including photoionization by both stellar and
non-stellar continua, as well as excitation by shocks. For a review of
the properties of LINERs and the plausible excitation mechanisms, we
direct the reader to \citet{f96} and \citet{b02}. Although LINERs as a
class are heterogeneous, their continuity with Seyfert galaxies in
optical luminosity and relative line strengths suggests that many
LINERs may be genuine AGNs powered by accretion onto a supermassive
black hole.

The fact that photoionization by a dilute, hard, power law continuum
reproduces the low-ionization spectra seen in LINERs
\citep[e.g.,][]{fn83,hs83} lends support to this theory. Furthermore,
\citet{hfs_paper4} found that $\sim$ 20$\%$ of the LINERs in their
ground based survey exhibited broad H$\alpha$ emission lines, similar
to, but significantly less luminous, than those found in Seyfert 1
galaxies. Given the difficulty of detecting the broad H$\alpha$
emission in these galaxies, \citet{hfs_paper4} noted that many more
LINERs could have broad H$\alpha$ emission lines. Those LINERs which
emit broad H$\alpha$ lines show other AGN-like features, such as steep
spectrum radio and hard X-ray cores \citep{ho01}

In the X-ray and UV bands, WLRGs bear many striking resemblances to
those LINERs which are AGNs.  An ASCA survey of LINERs and other Low
Luminosity AGNs by \citet{terashima} showed that the X-ray
luminosities of LINERs were 2 -- 3 orders of magnitude smaller than
those of classical Seyfert galaxies. The authors did not find that the
X-ray spectra of LINERs were harder than those of more luminous
radio-quiet galaxies, however. In the UV, the spectral energy
distributions (SEDs) of accretion-powered LINERs show a significant
deficit relative to Seyfert galaxies \citep{ho99}. Although, it is
difficult to measure the SED for WLRGs in the optical-UV regime, a
similar decrement is seen in the WLRG 3C~270 \citep[a.k.a. NGC
4261;][]{zb98,ho99}.

Thus there is mounting evidence that both the emission-line properties
and the underlying X-ray and UV ionizing continuum, are similar in
WLRGs and accretion-powered LINERs.  Should a connection between
accretion-powered LINERs and WLRGs be confirmed, this would have
important implications for the central engines of these objects. While
the central engine in Seyferts is thought to be fed by a thin
accretion disk, \citet{ho99} suggests that LINERs may be fed by
Advection Dominated Accretion Flows (ADAFs; Narayan \& Yi 1994, 1995)
or a similar structure. Since the UV photons generally created in the
inner regions of a thin accretion disk are not efficiently emitted in
an ADAF, this scenario explains the lack of a UV bump and the low
X-ray luminosity seen in LINERs. An ADAF could arise if the mass
accretion rate is substantially below the Eddington accretion rate.
{\it If} accretion-powered LINERs and WLRGs are both fueled by ADAFs,
any successful ADAF model must be able to explain the properties of
both systems, thereby placing additional constraints upon theoretical
models of the accretion flow. In particular, the central engine must
be able to produce kpc-sized jets in some objects but not others,
which is a significant constraint on theoretical models.

Motivated by the above issues we have undertaken a more detailed
spectroscopic study of WLRGs. The results of \citet{t98} merely
suggest that the emission-line properties of WLRGs are similar to
those of LINERs. Whether or not this is the case has not been
established because the emission-line ratios required for a robust
classification have not been measured. In \S\ref{data} we describe the
observations and data reductions. Then in \S\ref{meas} we present
optical spectra of 20 of the 26 WLRG from the \citet{t98} sample and
measure the necessary line ratios to test whether or not WLRGs satisfy
the H80 LINER criterion. We supplement our observations with data for
four additional WLRGs from the literature and include data for three
WLRGs in our sample for which the literature provides emission line
ratios not measurable from our spectra or which improve upon our
measured limits. We also classify WLRGs in the alternative
emission-line galaxy schemes of \citet{bpt} and \citet{vo87}. In
\S\ref{cloudy}, we perform photoionization calculations to determine
whether the observed emission-line ratios can be produced by the SED
of 3C~270, which we take to be representative of the SEDs of
WLRGs. We use these calculations to place constraints on the
ionization parameter and hydrogen density of the emission-line
gas. In \S\ref{discussion} we discuss the viability of an
accretion-powered central engine as the photoionizing source and
consider alternative excitation sources which may be present. Finally,
in \S\ref{conclusions}, we summarize our results and suggest avenues
for further work. Throughout this paper we assume a Hubble constant of
$H_0=70~{\rm km~s^{-1}~Mpc^{-1}}$ and a deceleration parameter of $q_0=0.5$.

\section{OBSERVATIONS AND BASIC DATA REDUCTIONS \label{data}}

The new data set used here consists of spectra of 20 WLRGs, as listed
in Table \ref{obs}. The observations were carried out at Kitt Peak
National Observatory (KPNO), MDM Observatory, Lick Observatory, and
Cerro Tololo Interamerican Observatory (CTIO) over a period of three
years, with dates as given in Table \ref{obs}. Spectra were typically
taken through $1.\!\!^{\prime\prime}5$--$1.\!\!^{\prime\prime}9$ slits
oriented at the parallactic angle whenever necessary. The typical
seeing was $1.\!\!^{\prime\prime}5$--$2.\!\!^{\prime\prime}0$. The
data were reduced in a standard fashion. In summary, preliminary
reductions included subtraction of the bias level and division by the
flat field. One-dimensional spectra were extracted from a
$0.\!\!^{\prime\prime}8$--$1.\!\!^{\prime\prime}3$ window centered on
the nucleus of each galaxy. Wavelength calibration was carried out
using arc spectra obtained immediately after the object exposure. Flux
calibration was carried out with the help of standard stars observed
on the same night and reduced in the same manner as the object. The
spectrum of the standard star was also used to derive a correction
template for discrete atmospheric absorption bands. The final spectral
resolution was 6--7$\rm \AA$ for spectra taken with the CTIO 1.5m and
Lick 3m telescopes and 3.5--4.5$\rm \AA$ for spectra taken with the
KPNO 2.1m and MDM 2.4m telescopes.

The observational details for the supplementary observations are
contained in Table \ref{supp_obs} and the references therein.

\section{EMISSION-LINE MEASUREMENTS AND CLASSIFICATION \label{meas}}

\subsection{Emission-Line Measurements}

Although the [\ion{O}{2}]$\lambda 3727$/[\ion{O}{3}]$\lambda 5007$ and
[\ion{O}{1}]$\lambda6300$/[\ion{O}{3}]$\lambda 5007$ line ratios are a
good diagnostic for discriminating between different types of emission
line galaxies and define LINERs as a class, they are not ideal from an
observational point of view. They are greatly affected by internal
reddening, which is generally unknown, and it can be quite
inconvenient to obtain measurements of these widely spaced line pairs.
To overcome these difficulties, \citet{vo87} and \citet{bpt} developed
additional 2-dimensional diagnostic diagrams which are also useful for
classifying emission-line galaxies, based upon the following pairs of
emission line ratios: [\ion{O}{1}]$ \lambda 6300$/H$\alpha$
vs. [\ion{O}{2}]$\lambda3727$/[\ion{O}{3}]$\lambda5007$; [\ion{O}{3}]$
\lambda 5007$/H$\beta$ vs.  [\ion{S}{2}]$ \lambda\lambda$
6716,6731/H$\alpha$; [\ion{N}{2}]$ \lambda 6583$/H$\alpha$
vs. [\ion{O}{2}]$ \lambda3727$/[\ion{O}{3}]$\lambda5007$;
[\ion{O}{3}]$ \lambda 5007$/H$\beta$ vs. [\ion{N}{2}]$ \lambda
6583$/H$\alpha$; [\ion{O}{3}]$ \lambda 5007$/H$\beta$ vs. [\ion{O}{1}]
$ \lambda 6300$/H$\alpha$.  These diagnostic diagrams are now widely
used in addition to the H80 criterion to differentiate between LINERs
and other narrow emission-line galaxies, such as Seyfert 2s and
\ion{H}{2} regions \citep[e.g.;][]{sf90,ft92}.

The spectrum of each object contained the combined starlight of the
host galaxy in addition to the low-ionization emission-lines we would
like to study. To accurately determine the line strengths, a template
galaxy spectrum was used to subtract the starlight contribution.  As a
first step to subtracting the template galaxy spectrum, the redshift
of each WLRG was determined using strong narrow emission lines or
prominent stellar absorption lines. In general, the spectra possessed
strong emission lines at both their blue and red ends, allowing for an
accurate redshift determination. When no clear blue emission lines
were present, resolved \ion{Ca}{2} $\lambda \lambda 3934,3968$ K and
H) absorption lines from the host galaxy were used to increase the
number of measurements. In a few instances the spectrum contained no
strong emission or absorption lines and the redshift was obtained from
\citet{t98}. Table \ref{info} contains a list of redshifts with errors
(column~2) as well as the number of lines used to obtain this redshift
(column~3).

\begin{figure}
\epsscale{1.0}
\plotone{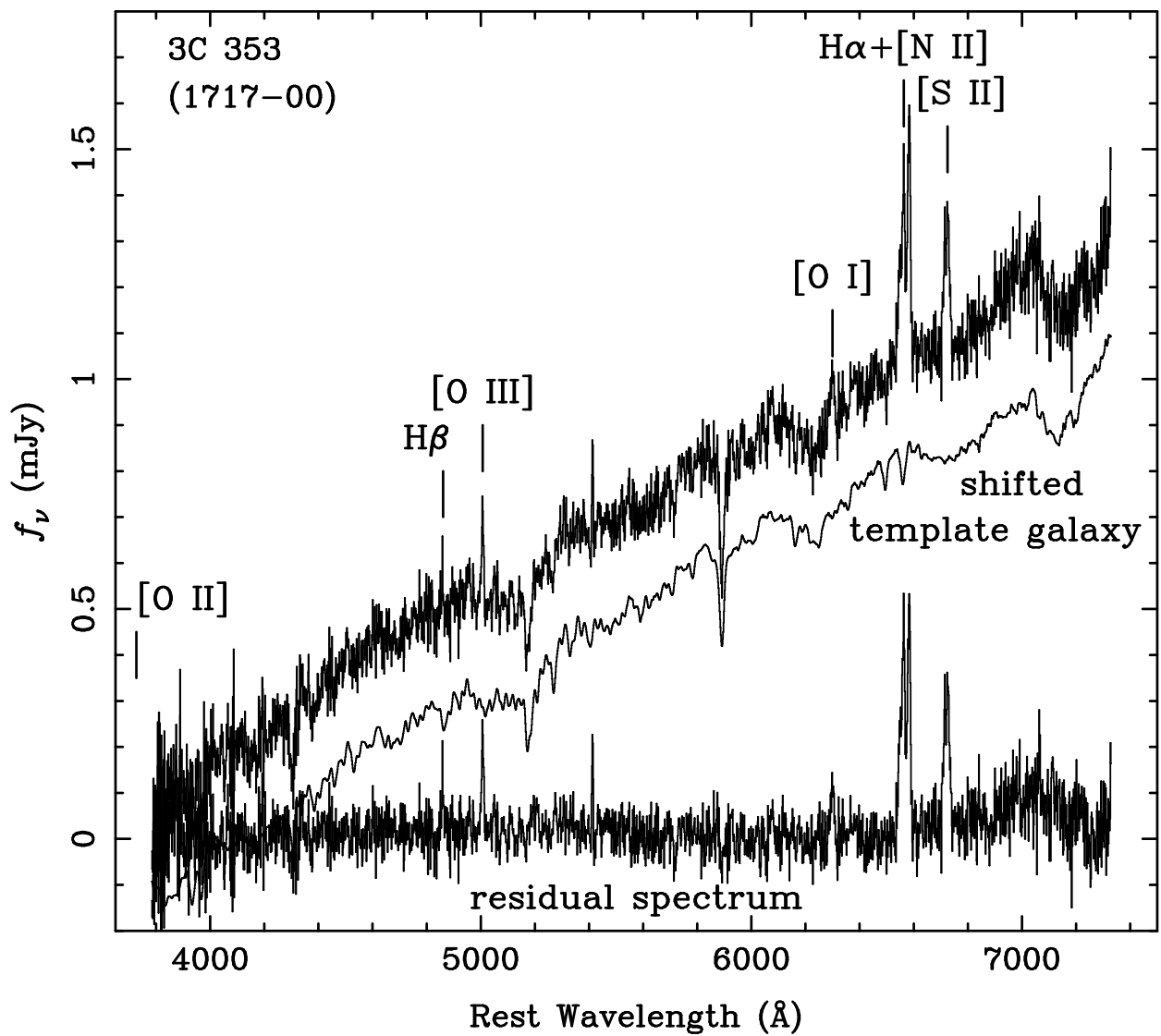}
\caption{\label{1717} Example of galaxy template subtraction 
for 1717--00 (3C 353), which is an average spectrum in terms of initial
line strength, S/N, and resolution. The uppermost spectrum is the raw
spectrum of the galaxy, which has been de-reddened and shifted to the
rest frame of the template galaxy.  In the middle is the template
galaxy and the residual spectrum is shown along the bottom of the
plot.}
\end{figure}

We modeled the continuum with a starlight template and a featureless
power law component when necessary. Occasionally, an additional
low-order polynomial was required to adequately model the
continuum. The spectra of giant ellipticals, which host all known
WLRGs, are extremely uniform.  Therefore, our template library
consisted of only four objects (NGC~1399, 3379, 4339, 4365, 5332) with
Hubble types ranging from S0 to E3.  After correcting each WLRG
spectrum for galactic reddening (listed in column 4 of
Table~\ref{info}) and shifting the spectrum to the rest frame, the
continuum was determined with a $\chi^{2}$ minimization routine using
four fitting regions in which the observed flux was dominated by
emission from the galaxy. The best-fit continuum was then subtracted
and the residuals visually inspected to ensure that the subtraction
routine did not leave artificial features in the residuals. In
Fig. \ref{1717}, we show the result of the starlight subtraction for
3C 353 (1717-00), a typical galaxy from our sample in terms of S/N and
resolution.

Isolated emission lines were fitted with Gaussians with the wavelength
and FWHM of the line as free parameters. The closely-spaced
[\ion{S}{2}]$\lambda\lambda 6716,6725$ doublet was fitted using a
de-blending algorithm, in which the FWHM of the lines were
independent. The [\ion{N}{2}]$\lambda\lambda 6548,6583$ and H$\alpha$
lines were severely blended.  Therefore, we used a combination of
Gaussians to fit [\ion{N}{2}]$\lambda\lambda 6548,6583$ in a 1:3 ratio
with the FWHM of the two lines forced to be the same. The fit to the
[\ion{N}{2}] lines was determined primarily by the stronger $\lambda
6583$ line which was sufficiently well separated from H$\alpha$ line
that its peak was clearly seen. The residual H$\alpha$ line was then
fitted with a simple Gaussian.  In all cases, the best fit was
determined using a $\chi^{2}$ minimization routine. No evidence of
broad H$\alpha$ is seen, although it would be difficult to detect, as
discussed by \citet{hfs_paper4}, particularly in low S/N data. In
those cases in which no clear line was visible, we determined a
``1$\sigma$'' upper limit on the line flux by fitting the local noise
with a Gaussian with the peak fixed to the expected wavelength. The
FWHM was restricted to the range of the measured FWHM of similar lines
in the spectrum (i.e. forbidden or Balmer lines). Since the {\it
absolute} flux calibration was not necessarily the same for two
separate spectra of the same object, line ratios were only calculated
when both lines were measured from the same spectrum. The
emission-line ratios are reported in Table \ref{lines}.

\subsection{Assessment of Uncertainties \label{errors}}

There were several sources of error that contributed to an overall
error of 20--30$\%$ for each line-strength measurement. Using, 1717--00
as our typical WLRG we performed several sets of measurements and
found that the majority of the error is incurred in the measurement of
the flux, although the starlight subtraction contributes as well, as
discussed below. The fluxes of clean, strong lines were assigned a
20\% while low S/N or distorted lines were given a 30\% error.

We performed an extensive literature search in order to compare our
emission line ratios with those obtained by other authors. In total we
were found 23 line ratios with at least one measurement from the
literature. Assuming a 20\% error on the literature values, there were
discrepancies in only six instances. In 1717-00, five of the line
ratios had comparison values and only one was discrepant.  As
described in \S\ref{classification}, our classification system is
fairly rigorous and a classification error is only likely to occur if
several ratios from the same object are incorrect. Given the results
of the literature search, this is unlikely to occur.

Although most of the error arose from the flux measurement, the galaxy
subtraction introduced some systematic errors, particularly in the
[\ion{O}{3}]--H$\beta$ and [\ion{N}{2}]--H$\alpha$ regions in which
broad H$\beta$ or H$\alpha$ absorption is expected. In particular, a
mismatch between the stellar populations in a WLRG and the template
galaxy would lead to an improper subtraction of the underlying Balmer
absorption and the fluxes of the H$\alpha$ and H$\beta$ emission lines
would be either under- or overestimated in that object. In a few
instances, the [\ion{N}{2}]$\lambda 6548$/[\ion{N}{2}]$\lambda 6583$
ratio was not 1/3, as expected from atomic physics, indicating that
the H$\alpha$ absorption line may have in fact been improperly
subtracted. There are four objects in which the H$\alpha$/H$\beta$
ratio, listed in column 8 of Table~\ref{lines}, was considerably {\it
lower} than expected from case B recombination: 0034--01, 1246--41,
2058--28, and 2104--17. This suggests that the H$\alpha$ flux was
underestimated or the H$\beta$ flux overestimated.  The
[\ion{O}{3}]/H$\beta$ ratios for 0034--01 and 2058--28 obtained from
the literature are considerably higher than we measured (see
Table~\ref{lines} and its footnotes), indicating that our measured
H$\beta$ flux may indeed be too large by almost a factor of
two. Therefore, we treat the measured H$\beta$ fluxes of these objects
and the ratios in which they appear with caution.

\begin{figure}
\epsscale{1}
\plotone{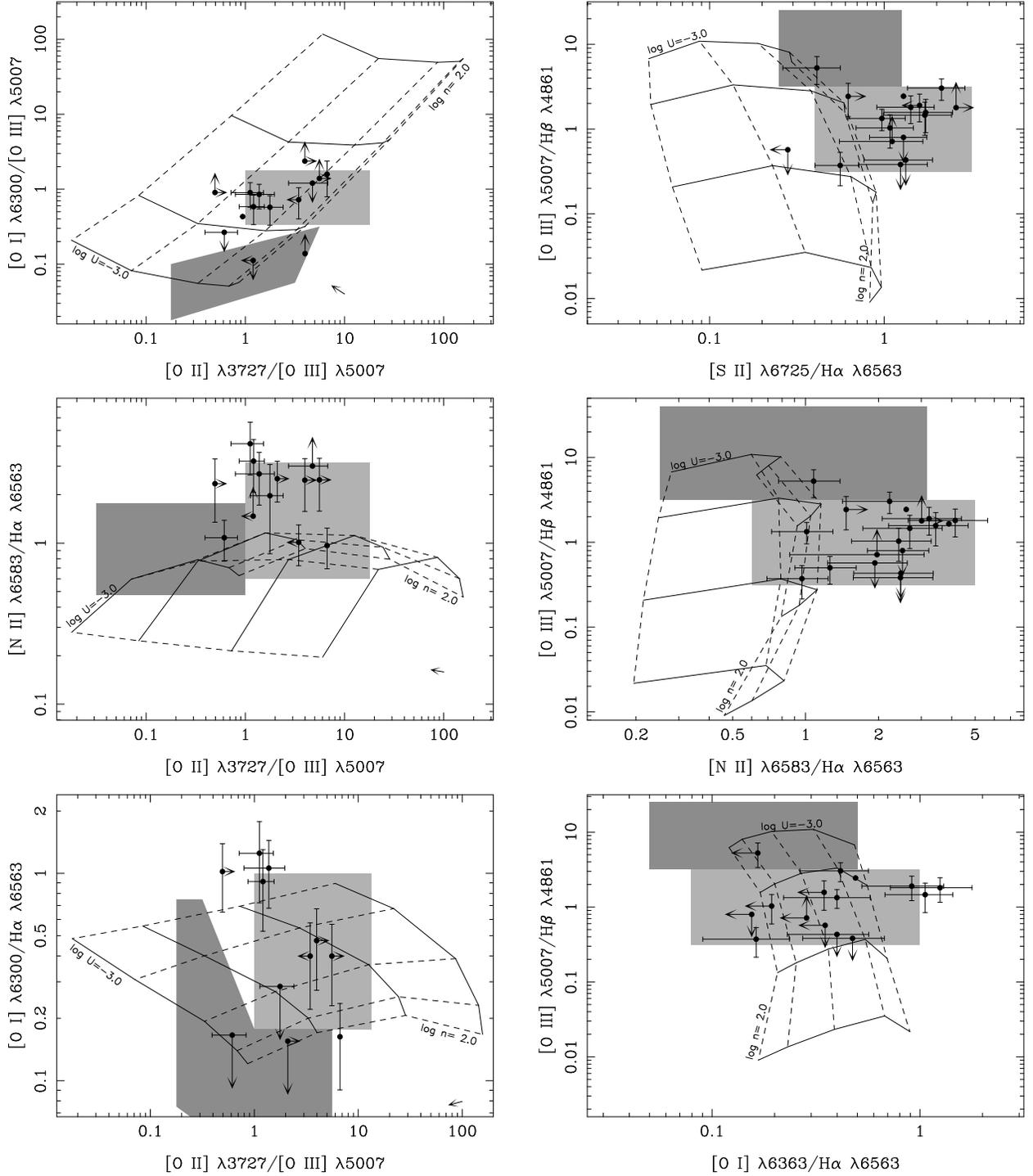}
\caption{\label{BPT} Classification of 22 WLRGs in the 2D diagnostic 
  diagrams.  The regions typically occupied by LINERs are shown in
  light grey while those occupied by Seyfert 2s are dark grey. The
  results of our photoionization calculations are overlaid. The
  Ionization Parameter ($U$) varies from $10^{-3}$ to $10^{-4.5}$, with lines
  of constant $U$ represented with solid lines. The Hydrogen density ($n$)
  varies from $10^{2}$ to $10^{6}~{\rm cm}^{-3}$, with lines of constant 
  $n$ shown as dashed lines.}
\end{figure}

Another source of uncertainty, independent of the above measurements,
was the internal reddening, which we have ignored. Reddening vectors,
assuming one magnitude of visual extinction, are included in all of
the diagnostics diagrams involving the widely spaced oxygen ratios
(Fig. \ref{BPT}).  One diagnostic of the reddening is the
H$\alpha$/H$\beta$ ratio, listed in col. 8 of Table \ref{lines}.
H$\alpha$/H$\beta$ is expected to be 3.1 for AGN models \citep{hs83}
or 2.86 for H II regions and planetary nebulae \citep{b71}. Of the 14
objects with measurements or limits of H$\alpha$/H$\beta$ only 0131-36
and 0625-53 appear to be significantly reddened. An enlarged
H$\alpha$/H$\beta$ ratio could also result if the WLRG has a younger
stellar population, and hence deeper Balmer absorption lines, than the
template galaxy. Although the fluxes of both H$\beta$ and H$\alpha$
would be underestimated, the effect would be proportionately more
severe in the weaker H$\beta$ line.  This could be the case for
0131-36; H$\gamma$ and H$\delta$ are seen in absorption \citep{t93}
indicating the presence of hot stars and the soft X-ray spectrum is
consistent with the presence of a starburst \citep{s99}, perhaps
located in the prominent dust lane \citep{ws66}.

A final concern was that the emission line ratios were sensitive to the
physical size of the extraction region (cols 6 and 7 of Table
\ref{info}).  We chose as small an extraction region as
possible, so that the emission arose primarily from the
true ``Narrow Line Region'' (NLR; i.e. that which is excited by the
nucleus). However, the extraction aperture in the higher-redshift
objects encompassed a region that extends well past the true NLR and
could include additional emission-line sources. Additionally, in some
of the nearby objects, the extraction region was less than 500 pc in
size and perhaps some of the NLR was excluded. To test this, we
extracted spectra from 0320--37 and 0131--36 using
$2.\!\!^{\prime\prime}6$, $3.\!\!^{\prime\prime}9$, and
$6.\!\!^{\prime\prime}5$ extraction lengths, corresponding to physical
dimensions of 0.27, 0.41, and 0.68 kpc for the former object and 1.53,
2.29, and 3.83 kpc for the later. In both objects, the emission line
fluxes increased as the size of the extraction length increased,
although more significantly for 0320--37, but the ratios remained the
same within errors. Although each object must be treated individually,
these tests confirm that the emission line ratios were not extremely
sensitive to the size of the extraction region. In particular, it is
unlikely that the NLR is resolved or that additional emission sources
between 1.5 and 4 kpc produce significant contamination. In
\S\ref{discussion}, we discuss the possibility that contaminating
sources are contained even within the smallest extraction box.

Long-slit spectroscopy and narrowband imaging also suggest that there
is little contamination. \citet{baum92} performed a long-slit
spectroscopic survey of emission-line nebulae in radio galaxies,
including the WLRGs 0055--01, 0305+03, 0325+02, and 0915--11. Although
the authors found variations in the emission-line ratios within
individual objects, both with distance from the nucleus and position
angle, the variations from source to source were much
larger. Narrowband [\ion{N}{2}]+ H$\alpha$ images of the same WLRGs,
plus 0123--01, 0255+05, 1251--12, and 1717--00, showed that more than
90\% of the flux was contained within 2.68 kpc of the nucleus in most
objects \citep[adjusted using our assumed H$_0$,][]{baum89}. Even in
0915--11, which is known to have a circumnuclear ring of star
formation \citep{Mc95,Mel97}, 80\% of the light was contained within
2.68~kpc. \citet{Hutchings} found no emission outside of a
$0.\!\!^{\prime\prime}35 \times 0.\!\!^{\prime\prime}5$ box in a
spatially resolved spectrum of 0305+03 obtained with the {\it Hubble
Space Telescope.}  Therefore, for the purposes of classification and
comparison with photoionization models, we assume that the line ratios
that we measured were those of the NLR. We will return to this issue
in our discussion of the results in \S\ref{discussion}.

\subsection{Classification \label{classification}}

In Fig. \ref{BPT} we show the resulting 2-dimensional diagnostic
diagram used originally by H80 to classify LINERs, in addition to the
diagnostics developed by \citet{bpt} and \citet{vo87}. We include
medium and light shaded polygons that indicate the regions occupied by
a large concentration of Seyfert 2s and LINERs, respectively, based
upon the data given in \citet{bpt} and \citet{vo87}. The area occupied
by \ion{H}{2} regions has been omitted for clarity, as few WLRGs fall
in this portion of the diagram. There is some overlap between the
various emission-line galaxy classifications and we have selected
these regions only to indicate where the majority of objects from
these two classes are located. Extreme outliers do exist which would
be mis-classified by these diagrams; these regions only indicate where
the {\it typical} Seyfert~2 galaxies and LINERs are located.

We use all six diagnostic diagrams to place each object into one of
the following categories: LINER (L); Possible LINER (PL); Non-LINER
(NL) Conflicting (C); and No Classification (NC). The classification
for each object is given in the last column of Table \ref{lines}.  To
obtain this classification, we first divided the objects into three
groups for each diagram: LINER; Possible LINER; and Non-LINER. An
object was a LINER if the measured line ratios placed it in the LINER
region or the limits allowed the object to occupy the LINER region but
ruled-out both a Seyfert 2 and \ion{H}{2} region classification.
Occasionally, an object was clearly not a Seyfert 2 or \ion{H}{2}
region, but lay outside the LINER region. If outlying LINERs were
observed by \citet{vo87} or \citet{bpt} in this region of the diagram,
the object was also considered a LINER. When an object straddled two
zones or the limits allowed the object to be classified as a LINER and
a Seyfert 2 or \ion{H}{2} region, the object was labeled as a possible
LINER. Finally, an object which fell clearly into either the Seyfert 2
or \ion{H}{2} areas of the diagram or whose limits clearly excluded
the LINER region was placed in the Non-LINER group.

To arrive at the final classification, this grouping procedure was
performed for all six diagrams. An object was classified as a LINER
only when it qualified as a LINER in {\it all} the diagrams for which
the measurements existed. Likewise, an object was a Non-LINER only if it
was classified as such in all the diagrams in which it
appeared. Objects with a mixture of LINER and Possible LINER groupings,
were given the status of Possible LINER, while those objects which were
sometimes Non-Liners and other times LINERs or Possible LINERs were
classified as Conflicting objects. 

Only in 14 of the 20 objects were we able to apply the original H80
classification scheme because the [\ion{O}{2}] line was not observed
or both emission lines in one of the pairs were upper limits. Of
these, 8 are classified as LINERs, 4 are Possible LINERs, and 2 are
Non-LINERs. However, when the alternative classification schemes
developed by \citet{vo87} and \citet{bpt} are used, every object
except 0123-01 and 1648+05 are classified in at least one
diagram. With the exception of 0055--01, 1251--12, 1637--37, and
2211--17, the remaining objects are classified in at least 3 diagrams,
allowing for a fairly robust classification. In total, there are 12
LINERs (50\%), 6 Possible LINERs (25\%), 1 Non-Liner (4\%), 3
Conflicting classifications (13\%) and 2 Not Classified (8\%).  The
breakdown between the 13 FR I and 9 FR II galaxies which have been
classified is as follows: FR I -- 8 LINERs (61.5\%), 3 Possible LINERs
(23.1 \%), and 2 Conflicting (15.4 \%); FR II - 4 LINERs (44.5\%), 3
Possible LINERs (33.3\%), 1 Non-LINER (11.1\%), and 1 Conflicting
(11.1 \%).

With improved S/N data, several Possible LINERs may be established as
LINERs, since many of the objects were relegated to the Possible LINER
category because only a limit was obtained for one of the line ratios.

\section{PHOTOIONIZATION CALCULATIONS \label{cloudy}}

\begin{figure}[ht]
\epsscale{1}
\plotone{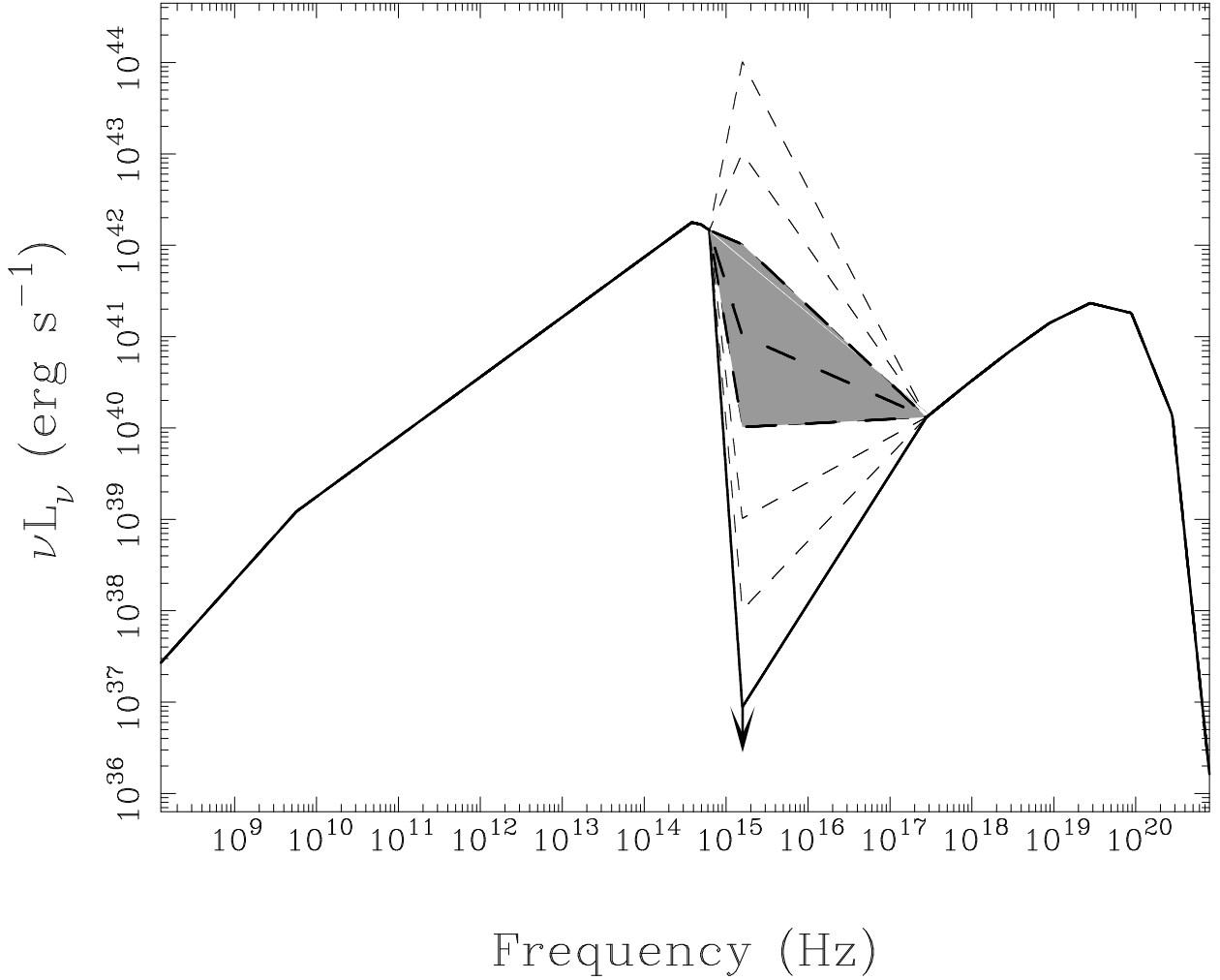}
\caption{\label{sed}The observed spectral energy distribution
of the WLRG 3C 270 (NGC 4261). This SED utilizes radio data from
\citet{t98}, optical data from \citet{ffj96}, UV upper limits from
\citet{zb98} and X-ray data from \citep{c02}. We have included an
exponential cut-off in the X-ray with a knee at 200 keV.  In our
photoionization calculations, we allow the UV flux to vary,
represented with the dashed lines, as discussed in \S\ref{cloudy}}
\end{figure}

In the previous section, we demonstrated that the vast majority of
WLRGs are classified as LINERs or Possible LINERs. Thus the suggestion
that WLRGs are related to LINERs is quite reasonable. There is no
doubt that WLRGs contain AGNs, as evidenced by the kpc-sized radio
jets and the central engine of the AGN is certain to play some role in
photoionizing the line-emitting region. However, it remains to be
determined whether the SED of a WLRG (such as 3C 270, Fig. \ref{sed})
{\it alone} can excite the observed emission lines, as is postulated
for LINERs, or whether additional sources, such as shocks or stellar
photoionization, are required. Furthermore, if WLRGs and LINERs are
related objects and both are photoionized by their central engines,
one might expect the properties of the line emitting gas, such as the
ionization parameter and gas density, to be similar to those inferred
for LINERs.

To address these questions, we performed photoionization calculations
using Cloudy 94.0 \citep{Ferland}. We used the measured SED of 3C 270,
using the data points of \citet{t98} (radio), \citet{ffj96} (optical),
\citet{zb98} (UV), and \citet{c02} (X-ray). We used a power law
interpolation between these data points and extended the X-ray portion
of the SED using the 2--10 keV power law with a high energy exponential
cutoff at 200 keV. In Fig. \ref{sed}, we plot the adopted SED of 3C
270.

The SED of 3C 270 shows an enormous deficit in the UV flux, while the
SEDs of LINERs generally exhibit a more moderate deficit
\citep{ho99}. \citet{J93} found a disk of gas and dust surrounding the
nucleus of the galaxy, which was clearly seen in the H$\alpha$ image
obtained with the {\it Hubble Space Telescope} \citep{M99}. This disk
may obscure the central engine, particularly in the UV.  It is thus
possible that the severe deficit of UV emission seen in 3C 270 is
primarily the result of obscuration by the dust disk and that the UV
flux seen by the line-emitting gas may be substantially higher than
what is observed. To explore this, we allowed the UV data point to
vary, as shown by the dashed lines in Fig. \ref{sed}.  We constructed
eight SEDs which, beginning with the measured SED, have
optical-to-xray spectral indices of $\alpha_{\rm ox}$ = 0.0, 0.3, 0.7,
1.0, 1.3, 1.7, 2.0, and 2.3, where $$ f_{\nu} \propto
\nu^{-\alpha_{\rm ox}}
\qquad {\rm and}\qquad
\alpha_{\rm ox} = 1+
{\left[\log(\nu f_{\nu})\vbox to 0.9em{\vss}\right]_{\rm 2500~\AA} - 
\left[\log(\nu f_{\nu})\vbox to 0.9em{\vss}\right]_{\rm 2~keV}
\over 2.61}\; .
$$

Using each of the possible UV fluxes shown in Fig. \ref{sed}, we
computed a photoionization grid in the hydrogen density ($n_{\rm H}$ =
10$^{2}$, 10$^{3}$, 10$^{4}$, 10$^{5}$, 10$^{6}$ cm$^{-3}$) and
ionization parameter ($U$ = 10$^{-3.0}$, 10$^{-3.5}$, 10$^{-4.0}$,
10$^{-4.5}$). The line emitting gas was assumed to have a solar
abundance pattern and metallicity and we ignored the effects of dust
grains, although they could survive. The line-emitting gas had a
plane-parallel geometry and the hydrogen density was held constant
throughout. The gas was radiation-bounded and the calculation ended
when the temperature of the gas reached 4000~K.  When the dielectronic
recombination rates were unknown, which is the case for most elements
in the 3rd and 4th rows of the periodic table, the mean of the rates
for C, O, and N were used for the four lowest stages of ionization.

We find that the {\it observed} SED ($\alpha_{\rm ox}$ = 0.0) of 3C
270 cannot reproduce either the data points for WLRGs or the LINER
region in the diagnostic diagrams; the UV flux seen by the
emission-line gas must actually be within the grey-shaded region
indicated in Fig. \ref{sed} ($\alpha_{\rm ox}$ = 1.0 -- 1.6), making
the inferred SED of 3C 270 more similar to those of LINERs
\citep{ho99}. We have overplotted the data grids obtained using the
SED with $\alpha_{\rm ox} = 1.3$, which produced the best agreement
between the model grids and both the data and the conventional LINER
region in all six diagnostic diagrams. The model points calculated by
Cloudy (assuming $\alpha_{\rm ox}=1.3$) at each $U$ and $n$, for all
six line ratios are listed in Table \ref{cloudygrid}. With the
exception of the [\ion{N}{2}]/H$\alpha$ and [\ion{S}{2}]/H$\alpha$
ratios, the modified SED of 3C 270 is quite successful at reproducing
the observed emission-line ratios of the WLRGs in our sample.

From these photoionization grids, we see that a wide range in hydrogen
density is required to reproduce all of the data points and to cover
the LINER region, similar to what is seen in LINERs
\citep[e.g.,][]{hfs93}. The ionization parameter, $U$, is typically
between 10$^{-3.5}$ and 10$^{-4.0}$, again, similar to that inferred
for LINERs \citep[e.g.,][]{hfs93}. It appears that not only are WLRGs often
classified as LINERs, but that the emission-line properties of WLRGs
are consistent with photo-ionization by an SED which is quite similar
to that seen in LINERs.

\section{DISCUSSION \label{discussion}}

Although it seems fairly plausible that the underlying excitation
mechanisms in LINERs and WLRGs are related, there are still several
outstanding questions which must be addressed. First, we must
demonstrate that the modified SED of 3C 270 produces enough ionizing
radiation to generate the luminosities of the observed
emission-lines. Second, if the emission-line gas in WLRGs is in fact
photoionized by the modified SED of 3C 270, a significant amount of
extinction in the UV along our line of sight to the nucleus is
implied; is this reasonable? Third, we must consider the large
discrepancy between the observed and predicted [\ion{N}{2}]/H$\alpha$
and [\ion{S}{2}]/H$\alpha$ ratios. Heating by cosmic rays may be
responsible for the enhancement of [\ion{N}{2}]/H$\alpha$ and
[\ion{S}{2}]/H$\alpha$, but it might also be caused by off-nuclear
emission sources within the extraction region. Finally, simply because
the SED of 3C 270 {\it can} reproduce most of the emission line ratios
does not imply that this {\it is} the actual mechanism; alternative
excitation mechanisms must be explored.

One way to determine whether the modified SED of 3C 270 is capable of
powering the emission lines is to calculate the photon budget for
H$\beta$. \citet{hfs_paper4} report that the reddening-corrected
H$\alpha$ flux is $1.51\times 10^{-14}~{\rm erg~ s^{-1}~
cm}^{-2}$. These authors assumed that the intrinsic values of the
H$\alpha$/H$\beta$ ratio is 3.1, thus the H$\beta$ luminosity is
$1.4\times 10^{37}~{\rm erg~s}^{-1}$ and the emission rate of
H$\beta$ photons, is $Q_{\rm H\beta}=3.40\times 10^{48}$ s$^{-1}$.
For Case B recombination, $Q_{\rm H\beta} = 0.12\; Q_{\rm ion}
f$ \citep{AGN2}, where $Q_{\rm ion}$ is the ionizing photon rate and
$f$ is the covering factor. Using the best-fit (modified) SED
described in \S\ref{cloudy}, we obtain $Q_{\rm ion} = 2.24\times
10^{51} {\rm s}^{-1}$. We see that even if the covering factor is as
small as 0.012, the AGN is capable of producing the observed emission
line luminosities.

If we wish to rely upon photoionization by the central AGN to produce
the observed emission-line ratios, then the UV flux seen by the
emission-line gas must be substantially larger than is measured,
requiring 8 -- 13 magnitudes of extinction (for $\alpha_{\rm ox}$ =
1.0 -- 1.6) at 2200\AA. Using the galactic extinction curve of
\citet{seaton} and A$_{\rm V}$/E(B-V) = 3.2, we find that A$_{\rm V}$
= 2.5 -- 4.2 magnitudes, assuming that the true UV flux is {\it at}
the upper limit. Although 3C 270 has an unresolved nucleus in the
optical and near-infrared {\it HST} images, the nucleus is not
detected in the UV by either \citet{zb98} or more recently by
\citet{a02}. Instead, the UV {\it HST} images of 3C 270 are dominated
by a dark disk, about 300 pc in size \citep{J93, a02}. \citet{c02}
postulate that this dusty disk may be responsible for the absorption
of the nuclear UV flux.  These authors estimate that this disk could
produce 1.2--2.5 magnitudes of visual extinction. This estimate was
made by assuming that 1/2 (for A$_{\rm V} = 1.2$) or 1/3 (for A$_{\rm
V} = 2.5$) of the starlight is produced behind the disk and then
comparing the brightness of the galaxy inside and outside the
projection of the disk. If there is emission from the disk itself,
A$_{\rm V}$ could be larger. The ratio of H$\alpha$/H$\beta$ measured
by \citet{hfs_paper2} also implies a low A$_{\rm V}$ of only 1.4
magnitudes. While the dusty disk is sufficient to produce the observed
extinction to the optical emission lines, it is just barely sufficient
to produce the minimum extinction necessary in the UV. It is possible
that dust is mixed with the narrow emission-line gas in which case the
extinction to the emission-line gas is less than to the UV
source. Alternatively, the covering factor of the dust could be larger
to the UV source than the emission-line gas. It is also possible that
in addition to the dusty disk, there is a different source of
extinction to the UV source.

From the {\it Chandra} and {\it XMM-Newton} observations of 3C 270,
column densities of $(6.2 \pm 0.9)\times 10^{22}$ and
$(4.3^{+1.8}_{-1.0})\times 10^{22} {\rm cm}^{-2}$ are inferred
\citep{c02,s03}.  Assuming a galactic gas-to-dust ratio, this 
implies $\sim$ 30 magnitudes of visual extinction to the X-ray
source. Observations of Seyfert galaxies suggest that it is not
unusual for the the visual extinction estimated from the X-ray data to
be 2 -- 10 times larger than estimated from optical and infra-red data
\citep{M01, G97}. Therefore, it is possible that the same medium
is responsible for the extinction of the X-ray and UV nuclear sources
as well as the optical emission-line region. 

On the other hand, as Sambruna et al. (2003) argue for 3C~270 (and
Weingartner \& Murray 2002 argue more generally referring to all AGNs)
the X-ray and optical/UV absorbers are likely to be distinct. The
optical/UV absorber is likely to be a dusty medium at a moderately
large distance from the AGN, while the X-ray absorber is likely to lie
closer to the AGN central engine and consist of neutral or ionized
gas, which is relatively free of dust and could have the form of an
outflowing wind. Either scenario should fulfill the following
requirements: the optical/UV extinction towards the nucleus/central
engine of 3C 270 should be higher than the extinction towards the
line-emitting region but considerably lower than what one would infer
from the X-ray extinction assuming a Galactic dust-to-gas ratio.

As seen in Fig. \ref{BPT}, even the modified SED of 3C 270 cannot
reproduce the large observed [\ion{N}{2}]/H$\alpha$ and
[\ion{S}{2}]/H$\alpha$ ratios in WLRGs. One could suppose the
H$\alpha$ is underestimated, but then one would expect the
[\ion{O}{1}]/H$\alpha$ to also be higher than expected in every object
with large [\ion{N}{2}]/H$\alpha$ and [\ion{S}{2}]/H$\alpha$ ratios,
which is not the case. However, several authors have encountered a
similar problem in modeling LINERs with photoionization models
\citep[e.g.,][]{hfs93,baum92}.  By artificially enhancing the nitrogen
abundance to 4 times solar, our photoionization grids reproduce the
[\ion{N}{2}]/H$\alpha$ ratio fairly well. Even when the sulfur
abundance is increased to 10 times solar, though, the large
[\ion{S}{2}]/H$\alpha$ ratio cannot be obtained.

As discussed by \citet{fm84}, cosmic ray heating can significantly
alter the ionization structure of the emission-line nebula leading to
enhancement of the low ionization lines. A low rate of cosmic-ray
heating leads to an increase in the [\ion{O}{1}]/H$\alpha$,
[\ion{N}{2}]/H$\alpha$ and [\ion{S}{2}]/H$\alpha$ ratios. Those
objects with unusually large [\ion{O}{1}]/H$\alpha$ ratios do tend to
also have large [\ion{N}{2}]/H$\alpha$ and [\ion{S}{2}]/H$\alpha$
ratios, however we have few measurements of [\ion{O}{1}]/H$\alpha$ in
our sample. At higher heating rates, however, the partially ionized
zone is destroyed, reducing the [\ion{O}{1}]/H$\alpha$ ratio while
increasing the [\ion{N}{2}]/H$\alpha$ and [\ion{S}{2}]/H$\alpha$
ratios by a factor of $\sim$ 2. Those objects with unusually small
[\ion{O}{1}]/H$\alpha$ do not have larger than average
[\ion{N}{2}]/H$\alpha$ and [\ion{S}{2}]/H$\alpha$, but these values
are larger than predicted by our photoionization calculation. Thus,
cosmic ray heating may be responsible for the enhancement in the
[\ion{N}{2}]/H$\alpha$ and [\ion{S}{2}]/H$\alpha$ ratios for some
objects. However, there are many objects with large
[\ion{N}{2}]/H$\alpha$ and [\ion{S}{2}]/H$\alpha$ ratios while having
normal [\ion{O}{1}]/H$\alpha$, which cannot be explained with cosmic
ray heating.

Another alternative would be to invoke additional off-nuclear sources
of [\ion{N}{2}] and [\ion{S}{2}] emission within the extraction
region. This is unlikely for several reasons. First, one would expect
that those objects with larger extraction regions would be more likely
to include these additional emission sources; this is not
seen. Second, as mentioned in \S\ref{errors}, the available narrowband
[\ion{N}{2}]+H$\alpha$ images show that the emission-line gas is
confined within 2.68 kpc \citep{baum89}, even for those objects with
large [\ion{N}{2}]+H$\alpha$. Furthermore, an inspection of {\it HST}
images \citep{baum88} shows that there are no clumps of emission
within this 2.68 kpc region. Finally, {\it HST} R-band
\citep{M99} and UV images \citep{zb98} of several WLRGs also show very
little structure.

Although it is unlikely that there are off-nuclear sources of
[\ion{N}{2}] and [\ion{S}{2}] emission, alternative nuclear excitation
mechanisms must be explored. \citet{ds95} find that shocks can
reproduce the range of emission-line ratios seen in Seyfert and LINERs
and are a good alternative to photoionization by an AGN. Shocks
certainly can be expected to be present in the nuclear regions of
radio galaxies, due to the interaction of the jets with ISM.  LINERs
are well modeled with fast shocks (150 -- 500 km s$^{-1}$) in a gas
poor environment while Seyferts require a richer environment, slightly
faster shocks (300 -- 500 km s$^{-1}$) as well as precursor UV
photons. 

\citet{d97} performed a detailed analysis of the optical and
UV emission lines in M87, a nearby, well-known LINER, hosted by a
radio galaxy. They conclude that the presence of strong UV emission
lines can only be explained by shock models. Furthermore, their
detailed modeling suggests that the outer accretion disk itself is a
plausible source of these shocks.  However, the best-fitting shock
model of \citet{d97} cannot reproduce the [\ion{N}{2}]/H$\alpha$ and
[\ion{S}{2}]/H$\alpha$ ratios without enhancing the N and S abundances
to 3 and 2 times solar, respectively. Even then, the
[\ion{S}{2}]/H$\alpha$ ratio is still underpredicted by $\sim$ 20\%,
which is substantially larger than the error reported on their
measurement. When we enhance our N and S abundances to those adopted
by \citet{d97}, we can reproduce the observed [\ion{N}{2}]/H$\alpha$
ratios as well as \citet{d97}, but we still underpredict the
[\ion{S}{2}]/H$\alpha$ ratio by $\sim$ 35\%.  Thus, the shock model
presented by \citet{d97} would be slightly better at reproducing the
large [\ion{S}{2}]/H$\alpha$ ratios observed in WLRGs. However in the
absence of UV data, this model is not clearly preferred over
photoionization.

Stellar photoionization by very hot, luminous Wolf-Rayet stars stars
has been postulated by \citet{tm85}, while photoionization by O stars
has been studied by \citet{ft92} and \citet{s92}. Like shocks, these
models can account for the observed emission-line ratios of WLRGs and
LINERs, but also fail to produce the large observed
[\ion{N}{2}]/H$\alpha$ and [\ion{S}{2}]/H$\alpha$ ratios without
enhancing the N and S abundances.  

Thus, photoionization by an AGN, as represented by the SED of 3C 270,
can reproduce the emission-line ratios presented here. Alternative
excitation mechanisms cannot be ruled out, but are not favored,
either. Since an AGN {\it must} be present to explain the radio jets
of WLRGs, economy of means suggests that the AGN should also be
responsible for powering the observed emission lines.  With the
current data, we are unable to determine conclusively which excitation
mechanism is responsible for the line-emission: photoionization by the
central engine; shocks between the jets and the ISM; or shocks within
a large accretion disk itself.

\section{CONCLUSIONS AND FUTURE PROSPECTS \label{conclusions}}

In this paper, we have presented emission line ratios for 24
Weak-Line Radio Galaxies (WLRG) and classified them according to
the H80 criterion, as well as the diagnostic diagrams developed by
\citet{vo87} and \citet{bpt}. We find that most of the WLRGs in our
sample are classified as either LINERs (50\%) or Possible LINERs
(25\%). One object, 0131--36, was found to be a Non-LINER and 3 objects
had conflicting classifications and may be hybrid objects. Only two
objects were not classified. With improved S/N measurements, it is
likely that many Possible LINERs may be promoted to LINERs status.

The measured emission-line ratios of WLRGs are reproduced fairly well
by photoionization by the SED of the WLRG 3C 270 when the UV flux is
increased substantially above the observed upper limit, yielding an
SED which is quite similar to those seen in LINERs \citep{ho99}. Thus
we conclude that our line of sight to the UV source must be obscured
by 2.5--4.2 magnitudes of visual extinction. The photoionization
calculations indicate that a wide range of hydrogen densities is
required to cover the data points, but that the ionization parameter
is typically between $10^{-3.5}$ and $10^{-4.0}$, as seen in LINERs.

The [\ion{N}{2}]/H$\alpha$ and [\ion{S}{2}]/H$\alpha$ ratios are much
larger than predicted by the photoionization grids, an effect noted by
several authors in the study of LINERs. The enhancement of these
ratios might be caused by cosmic ray heating, but is unlikely to be
caused by off-nuclear sources of emission. The emission-line ratios we
present here are also well re-produced by shock and stellar
photoionization models. However, the [\ion{N}{2}]/H$\alpha$ and
[\ion{S}{2}]/H$\alpha$ ratios are still under-predicted, so these
models are not favored over AGN photoionization based upon the data
presented here.

In conclusion, it is very likely that WLRGs and LINERs are intimately
related, not only in their emission-line properties but also in their
underlying excitation mechanisms. Higher S/N optical spectra of WLRGs
would help solidify the relationship between LINERs and WLRGs. More
importantly measurements of the weak [\ion{O}{3}]$\lambda 4363$ line
would help discriminate between photoionization and shock models, as
this line is expected to be much stronger in shock models. The most
progress can probably be made with narrowband and UV imaging and IR,
UV and long-slit spectroscopy. Although WLRGs are classified as LINERs
by their emission-line ratios, it is still uncertain whether
photoionization by an AGN is the only possibility. Through narrowband
and UV imaging with the {\it HST}, as well as long-slit spectroscopy,
valuable insight into distribution of the narrow emission-line gas
would be gained.  The observations of \citet{baum88}, \citet{baum90}
and \citet{zb98} have been extremely valuable in studying several
WLRGs in this sample, however each object is unique and must be
considered individually. UV and IR spectroscopy will be extremely
helpful in determining whether shocks or photoionization models are
favored as there are several additional diagnostics in these bands
(see Dopita et al. 1997; Filippenko 1996 and references therein).

\acknowledgements
We thank Jules Halpern and Joe Shields for assistance in obtaining
some of the spectra presented here, for assistance in carrying out and
interpreting the photoionization calculations, and for their critical
reading of the manuscript. We also thank the referee for insightful
comments. K.T.L. was supported in part by NASA grant NGT5-50387. R.M.S
acknowledges financial support form NASA LTSA grant NAG5-10708 and
from the Clare Boothe Luce Program of the Henry Luce Foundation. The
publication of this paper was made possible by a grant from the
Zaccheus Daniel Foundation. K.T.L. also acknowledges the generous
support of the NASA Pennsylvania Space Grant Consortium.

\clearpage

\begin{deluxetable}{lllccll}
\tabletypesize{\footnotesize}
\tablecaption{Journal of Observations \label{obs}} 
  \tablewidth{7.0in} 
  \tablehead{ 
    \colhead{IAU}  & \colhead{Other}   & \colhead{}          & \colhead{Exposure} & \colhead{Slit}         & \colhead{Spectral}    & \colhead{Observation} \\
    \colhead{Name} & \colhead{Name(s)} & \colhead{Telescope} & \colhead{Time (s)} & \colhead{Width ($''$)} & \colhead{ Range (\AA)} & \colhead{Date} 
            } 
  \startdata
  0034--01 & 3C 15              & KPNO 2.1m     & 1800       & 1.8  & 3860 -- 7506 & 1999 Dec 02\\
  0043--42 &                    & CTIO 1.5m     & 3000       & 1.8  & 5531 -- 9004 & 1999 Nov 03\\
           &                    &               & 3000       &      & 3677 -- 7131 & 1999 Nov 04\\  
  0123--01 & 3C 40              & MDM 2.4m      & 1200       & 1.5  & 6245 -- 8319 & 1998 Dec 20\\
           &                    &               & 2100       &      & 4217 -- 6289 & 1998 Dec 22\\
  0131--36 & NGC 612            & CTIO 1.5m     & 1800       & 1.8  & 3640 -- 7043 & 2001 Jan 23\\ 
  0305+03  & 3C 78, NGC 1218    & Lick 3m       & 900        & 1.5  & 4460 -- 7234 & 1988 Sep 12\\
  0320--37 & For A, NGC 1316    & CTIO 1.5m     & 1800       & 1.8  & 3640 -- 7043 & 2001 Jan 23\\
  0325+02  & 3C 88              & MDM 2.4m      & 1200       & 1.5  & 6245 -- 8319 & 1998 Dec 20\\
           &                    &               & 1800       &      & 4217 -- 6289 & 1998 Dec 22\\
  0427--53 & IC 2080            & CTIO 1.5m     & 900        & 1.8  & 3677 -- 7131 & 1999 Nov 04\\
  0453--20 & NGC 1692, OF-289   & MDM 2.4m      & 1200       & 1.5  & 6245 -- 8319 & 1998 Dec 20\\ 
           &                    &               & 1800       &      & 4217 -- 6289 & 1998 Dec 22\\
  0625--35 & OH-342             & CTIO 1.5m     & 4800       & 1.8  & 3881 -- 7287 & 2001 Jan 24\\
  0625--53 &                    & CTIO 1.5m     & 2400       & 1.8  & 3640 -- 7043 & 2001 Jan 23\\
  0915--11 & Hyd A, 3C 218      & KPNO 2.1m     & 1800       & 1.8  & 3860 -- 7506 & 1999 Dec 03\\
  1246--41 & NGC 4696           & CTIO 1.5m     & 1800       & 1.8  & 3640 -- 7043 & 2001 Jan 23\\
  1251--12 & 3C 278             & MDM 2.4m      & 900        & 1.5  & 6245 -- 8319 & 1998 Dec 20\\
           &                    &               & 900        &      & 4217 -- 6289 & 1998 Dec 22\\
  1318--43 & NGC 5090           & CTIO 1.5m     & 1800       & 1.8  & 3640 -- 7043 & 2001 Jan 23\\
  1333--33 & IC 4296            & CTIO 1.5m     & 1800       & 1.8  & 3640 -- 7043 & 2001 Jan 23\\
  1717--00 & 3C 353             & KPNO 2.1m     & 3600       & 1.8  & 3897 -- 7553 & 2000 Jun 04\\
  2058--28 & NGC 6998, OW-297.1 & CTIO 1.5m     & 1800       & 1.8  & 5795 -- 7593 & 1999 Nov 02\\
           &                    &               & 1800       &      & 3677 -- 7131 & 1999 Nov 05\\  
  2104--28 & NGC 7018, OX-208   & CTIO 1.5m     & 1800       & 1.8  & 5795 -- 7593 & 1999 Nov 02\\
           &                    &               & 1800       &      & 3677 -- 7131 & 1999 Nov 05\\  
  2211--17 & 3C 444             & KPNO 2.1m     & 3000       & 1.8  & 3860 -- 7506 & 1999 Dec 03\\
  \enddata   
\end{deluxetable}

\clearpage


\begin{deluxetable}{cllcclc} 
\tablecaption{Table of Supplementary Data from the Literature \label{supp_obs}} 
  \tablewidth{6.5in} 
  \tablehead{ 
    \colhead{IAU}  & \colhead{Other}   & \colhead{}          & \colhead{Exposure} & \colhead{Slit}       & \colhead{Observation} & \colhead{}\\
    \colhead{Name} & \colhead{Name(s)} & \colhead{Telescope} & \colhead{Time (s)} & \colhead{Width ($''$)} & \colhead{Date} & \colhead{Ref. \tablenotemark{a}}
            } 
  \startdata
0055--01  & 3C 29             & KPNO 4m  & 1800 & 2.0 & 1986 Nov & 1 \\
          &                   & ESO 2.2m & 1800 & 3.0 & 1990 Jul & 2 \\
0915--11  &                   & ESO 2.2m & 2400 & 2.0 & 1989 Mar & 2 \\
1216+06   & 3C 270, NGC 4261  & Hale 5m  & 1200 & 2.0 & 1986 Feb & 3 \\
1637--77  &                   & AAT 3.9m &  900 & 2.0 & 1992 Apr & 4 \\
          &                   & ESO 2.2m & 1200 & 2.5 & 1989 Mar & 2 \\
1648+05   & Her A             & ESO 3.6m & 1800 & 2.0 & 1989 Mar & 2 \\
1717--00  & 3C 353            & AAT 3.9m & 1900 & 2.0 & 1992 Apr & 4 \\
2211--17  &                   & ESO 3.6m & 1200 & 3.0 & 1990 Jul & 2 \\
\enddata
\tablenotetext{a}{{\it References --} (1) \citet{baum90}; (2) \citet{t93}; (3) \citet{hfs_paper2};
 (4) \citet{simp96}.}
\end{deluxetable}

\begin{deluxetable}{llcccccc}
\tabletypesize{\footnotesize}
\tablecaption{Basic Properties of WLRGs and Extraction Aperture for Their Spectra\label{info} } 
  \tablewidth{7.0in} 
  \tablehead{ 
            \colhead{IAU}   & \colhead{}                                    & \colhead{Num. of} & \colhead{} & \colhead{} & \colhead{Extraction} & \colhead{Extraction} & \colhead{} \\
            \colhead{Name}  & \colhead{\hspace{0.1cm}$z$ \tablenotemark{a}} & \colhead{Lines} & \colhead{E(B-V) \tablenotemark{b}}  & \colhead{FR Type \tablenotemark{c}} & \colhead{Width (kpc) \tablenotemark{d}} & \colhead{Length (kpc) \tablenotemark{d}} & \colhead{Ref. \tablenotemark{e}}
            } 
  \startdata
  0034--01  & 0.0733(6)               & 2 & 0.022 & II   & 2.37 & 2.00 & \\
  0043--42  & 0.1169(2)               & 5 & 0.011 & II   & 3.48 & 5.03 & \\
  0055--01  & 0.045 \tablenotemark{f} &   &       & I    & 1.72 & 0.72 & 1\\
            &                         &   &       &      & 2.58 & ?    & 2\\
  0123--01  & 0.019                   & 1 & 0.041 & II   & 0.54 & 1.05 & \\
  0131--36  & 0.0296(2)  	      & 5 & 0.020 & II   & 1.06 & 1.53 & \\
  0305+03   & 0.029      	      & 1 & 0.173 & I    & 0.85 & 0.42 & \\
  0320--37  & 0.0057(2)  	      & 3 & 0.021 & I    & 0.19 & 0.27 & \\
  0325+02   & 0.03051(8) 	      & 6 & 0.126 & II   & 0.88 & 1.93 & \\
  0427--53  & 0.0397(3)  	      & 3 & 0.009 & I    & 1.32 & 1.91 & \\
  0453--20  & 0.035      	      & 4 & 0.041 & I    & 1.02 & 1.98 & \\
  0625--35  & 0.055       	      & 1 & 0.067 & I    & 1.85 & 2.67 & \\
  0625--53  & 0.054 \tablenotemark{f} &   & 0.062 & II   & 1.82 & 2.63 & \\
  0915--11  & 0.05377(9) 	      & 4 & 0.042 & I    & 1.82 & 1.54 & \\
            &                         &   &       &      & 2.01 & ?    & 2 \\
  1216+06   & 0.006 \tablenotemark{f} &   &       & I    & 0.25 & 0.49 & 3 \\ 
  1246--41  & 0.0099(4)               & 5 & 0.113 & I    & 0.33 & 0.48 & \\
  1251--12  & 0.01574(1)              & 2 & 0.053 & I    & 0.45 & 1.22 & \\
  1318--43  & 0.0112(2)               & 4 & 0.144 & I    & 0.40 & 0.58 & \\
  1333--33  & 0.0126(4)  	      & 4 & 0.059 & I    & 0.47 & 0.68 & \\
  1637--77  & 0.041 \tablenotemark{f} &   &       & II   & 1.57 & ?    & 4\\
            &                         &   &       &      & 1.96 & ?    & 2\\
  1648+05   & 0.154 \tablenotemark{f} &   &       & I/II & 4.83 & ?    & 2\\
  1717--00  & 0.03040(7) 	      & 3 & 0.439 & II   & 1.09 & 0.92 &  \\
            &              	      &   &       &      & 1.19 & ?    & 4\\
  2058--28  & 0.0390(3)  	      & 2 & 0.113 & I    & 1.32 & 1.91 & \\
  2104--28  & 0.038      	      &   & 0.064 & II   & 1.29 & 1.86 & \\
  2211--17  & 0.153 \tablenotemark{f} &   & 0.115 & II   & 4.30 & 3.63 & \\
            &                         &   &       &      & 7.22 & ?    & 2
  \enddata 

  \tablenotetext{a\;}{The figure in parenthesis is the statistical
  uncertainty on the last digit. It was computed as the adjusted error in the 
  mean redshift: $S_{\rm n-1}=\sigma_{\rm n}/\sqrt{n-1}$, where $n$ is given in Col. 3. 
  No uncertainty is given when the redshift measurement was based on a single emission 
  line or taken from \citet{t98}.}
  \tablenotetext{b\;}{The Galactic color excess, taken from \cite{sch98}} 
  \tablenotetext{c\;}{Fanaroff-Riley Classification \citep{fr74} according to \citet{t98}.}
  \tablenotetext{d\;}{A question mark indicates that the extraction length was not specified.}
  \tablenotetext{e\;}{{\it References. --} (1) \citet{baum90}; (2) \citet{t93}; (3) \citet{hfs_paper2};
 (4) \citet{simp96}.}
  \tablenotetext{f\;}{Redshift obtained from \citet{t98}}

\end{deluxetable}

\clearpage

\begin{deluxetable}{clllllllll}
\tabletypesize{\footnotesize}
\tablewidth{6.8in}
\tablecaption{Measured Emission-Line Ratios \tablenotemark{a} \label{lines}}
\tablehead{
  \colhead{} & 
  \colhead{[\ion{O}{2}]} & 
  \colhead{[\ion{O}{1}]} &
  \colhead{[\ion{N}{2}]} & 
  \colhead{[\ion{O}{3}]} & 
  \colhead{[\ion{O}{1}]} & 
  \colhead{[\ion{S}{2}]} &
  \colhead{H$\alpha$} &
  \colhead{} &
  \colhead{} \\
\noalign{\vskip -8truept}
  \colhead{} & 
  \colhead{\hbox to 3 em{\hrulefill}} & 
  \colhead{\hbox to 3 em{\hrulefill}} &
  \colhead{\hbox to 3 em{\hrulefill}} & 
  \colhead{\hbox to 3 em{\hrulefill}} & 
  \colhead{\hbox to 3 em{\hrulefill}} & 
  \colhead{\hbox to 3 em{\hrulefill}} &
  \colhead{\hbox to 3 em{\hrulefill}} &
  \colhead{} &
  \colhead{} \\
\noalign{\vskip -4truept}
  \colhead{IAU Name} & 
  \colhead{[\ion{O}{3}]} & 
  \colhead{[\ion{O}{3}]} &
  \colhead{H$\alpha$} & 
  \colhead{H$\beta$} & 
  \colhead{H$\alpha$} & 
  \colhead{H$\alpha$} &
  \colhead{H$\beta$} &
  \colhead{Class.\tablenotemark{b}} &
  \colhead{Refs.\tablenotemark{c}}
          }
\startdata
0034--01 & $<$ 1.2        & $<$ 0.11     & $>$ 2        & 2(1)          &               & $>$ 0.6    & $<$ 2.3   &  C  &  \\
0043-42  & 1.2(3)         & 0.6(2)       & 3(1)         & 1.9(7)        & 0.9(4)        & $<$ 1.7    &           &  L  &  \\
0055--01 &                &              & 3.9\ft{d}    & 1.65\ft{d}    &               &            &           &  L  & 1,2 \\
0123--01 &                &              & $>$ 6        &               &               & $>$ 3      &           &  NC &  \\ 
0131--36 & 0.6(2)         & $<$ 0.27     & 1.1(3)       & 5(2)          & $<$ 0.16      & 0.4(2)     & 8(3)      &  NL &  \\
0305+03  &                & $<$ 0.6      & 3(1)         & 1.6(7)        & $<$ 0.3\ft{e} & 1.7(8)     & 2(1)      &  PL &  \\
0320--37 & 5(2)           & $<$ 1.2      & $>$ 3\ft{e}  & $>$ 2         &               & $>$ 3      &           &  PL &  \\
0325+02  &                &              & 2.2(8)       & 3.0(9)        & 0.4(2)        & 2.1(8)     &           &  L  &  \\
0427--53 & 1.8(7)         & 0.6(2)       & 2(1)         & $>$ 0.7       & $<$ 0.28      & 1.1(6)     & $>$ 2     &  PL &  \\
0453--20 &                &              & 2.4(9)       & 1.0(4)        & 0.14(6)       & 1.1(4)     &           &  L  &  \\
0625--35 & 1.1(4)         & 0.9(3)       & 4(1)         & 1.8(6)        & 1.3(5)        & 1.4(5)     & $>$ 0.7   &  L  &  \\
0625--53 & $>$ 0.5        & $>$ 0.9      & 2(1)         &               & 1.0(4)        &            & 7(3)      &  PL &  \\
0915--11 & 6.67\ft{f}     & 1.6(8)       & 1.0(3)       & 0.3(2)\ft{e}  & 0.20(7)       & 0.6(2)     & 4(1)      &  L & 2 \\ 
1216+06  &                &              & 2.60\ft{d}   & 2.44\ft{d}    & 0.49\ft{d}    & 1.29\ft{d} & 4.9\ft{d} &  L  & 3 \\
1246--41 & $>$ 6          & $>$ 1        & 2.5(9)       & $<$ 0.4       & 0.4(2)        & 1.3(6)     & 1.9(7)    &  L  &  \\
1251--12 &                &              & 1.3(4)       & 0.5(2)        &               &            &           &  L  &  \\
1318--43 & $>$ 2          &              & 2.5(7)       & $<$ 0.8       & $<$ 0.15      & 1.3(5)     & 3(1)      &  C  &  \\
1333--33 & $>$ 4          & $>$ 2        & 2.5(9)\ft{e} & $<$ 0.4       & 0.5(2)        & 1.2(5)     & 2(1)      &  L  &  \\
1637--77 & 0.94\ft{d}     & 0.43\ft{d}   &              & 9.61\ft{d}    &               &            &           &  L  & 2,4 \\
1648+05  & 4.55\ft{d}     &              &              & 0.62          &               &            &           &  NC & 2 \\
1717--00 & $<$ 3.45\ft{d} & 0.7(3)\ft{e} & 1.0(3)       & 1.3(4)        & 0.4(2)        & 1.0(4)     & 4(1)      &  PL & 4 \\
2058--28 &                &              & 1.9(7)       & $<$ 0.6\ft{e} & $<$ 0.4       & $<$ 0.28   & 1.3(6)    &  C  &  \\
2104--25 & 1.4(6)         & 0.9(3)       & 3(1)         & 1.5(6)        & 1.1(4)        & 1.7(7)     & 1.2(5)    &  L  &  \\
2211--17 & 4.0\ft{f}      & $>$ 0.3      &              & 0.57\ft{f}    &               &            &           &  PL & 2 \\
\enddata
\small
\tablenotetext{a}{ The figure in parenthesis is the uncertainty on the last 
digit. Values obtained from the literature do not have an uncertainty listed.}
\tablenotetext{b}{ The classification for each object, based upon the diagnostic diagrams 
as discussed in \S3 -- L = LINER, PL = Possible LINER, NL = Non-LINER, C = Conflicting, 
NC = Not Classified}
\tablenotetext{d}{ {\it References. --} (1) Baum et al. (1990); (2) \citet{t93}; (3) \citet{hfs_paper2};
 (4) \citet{simp96}.}
\tablenotetext{d}{ Taken from the literature.}
\tablenotetext{e}{ Our measurement is in conflict with reports in the literature, as discussed
in \S3.2 of the text. The values found in the literature are as follows:
 0305+03: [\ion{O}{1}]/H$\alpha$    = 0.50 (Baum et al. 1990); 
0320--37: [\ion{N}{2}]/H$\alpha$    = 2.13 \citep{p86}; 
0915--11: [\ion{O}{3}]/H$\beta$     = 0.75 \citep{t93}; 
1313--33: [\ion{N}{2}]/H$\alpha$    = 1.0  \citep{p86};
1717--00: [\ion{O}{1}]/[\ion{O}{3}] = 1.55 \citep{simp96}; 
2058--28: [\ion{O}{3}]/H$\beta$     = 1.03 \citep{t93}. 
}
\tablenotetext{f}{ Taken from the literature since our own spectra yielded only an upper or lower limit.}
\end{deluxetable}

\clearpage

\begin{deluxetable}{lrrrrrrrrrrr}
\tablenum{5}
\tablewidth{6.3in}
\tablecolumns{12}
\tablecaption{Line Ratios Predicted by Photoionization Calculations \tablenotemark{a}\label{cloudygrid}}
\tablehead{
\colhead{} &
\multicolumn{5}{c}{$\log (n/{\rm cm}^{-3})$} &
\colhead{} &
\multicolumn{5}{c}{$\log (n/{\rm cm}^{-3})$} \\
\noalign{\vskip -8pt}
\colhead{} &
\multicolumn{5}{c}{\hrulefill} &
\colhead{} &
\multicolumn{5}{c}{\hrulefill} \\
\colhead{\hbox to 3em{$\log U$}} &
\colhead{2} &
\colhead{3} &
\colhead{4} &
\colhead{5} &
\colhead{6} &
\colhead{\hbox to 2em{}} &
\colhead{2} &
\colhead{3} &
\colhead{4} &
\colhead{5} &
\colhead{6} 
}
\startdata
\sidehead{}
 & \multicolumn{5}{c}{[\ion{O}{2}]/[\ion{O}{3}]}&     & \multicolumn{5}{c}{[\ion{O}{1}]/[\ion{O}{3}]}   \\
\noalign{\vskip -8pt}
 & \multicolumn{5}{c}{\hrulefill}               &     & \multicolumn{5}{c}{\hrulefill}   \\
--3.0 & 0.864 &  0.684 & 0.332 &  0.071 & 0.018 &     & 0.057 &  0.051 & 0.055 &  0.081 & 0.207  \\
--3.5 & 4.023 &  3.311 & 1.598 &  0.331 & 0.084 &     & 0.320 &  0.288 & 0.279 &  0.347 & 0.822  \\
--4.0 & 28.49 &  24.44 & 12.60 &  2.729 & 0.727 &     & 4.603 &  4.160 & 3.877 &  4.250 & 9.599  \\
--4.5 & 157.1 &  144.1 & 86.44 &  21.97 & 6.003 &     & 55.45 &  50.85 & 49.43 &  55.51 & 117.3  \\
\sidehead{}
 & \multicolumn{5}{c}{[\ion{N}{2}]/H$\alpha$}   &     & \multicolumn{5}{c}{[\ion{O}{3}]/H$\beta$}   \\
\noalign{\vskip -8pt}
 & \multicolumn{5}{c}{\hrulefill}               &     & \multicolumn{5}{c}{\hrulefill}   \\
--3.0 & 0.628 &  0.703 & 0.789 &  0.597 & 0.279 &     & 6.210 &  8.033 & 10.21 &  10.87 & 6.755  \\
--3.5 & 0.924 &  1.046 & 1.157 &  0.773 & 0.247 &     & 1.565 &  2.034 & 2.822 &  3.322 & 1.948  \\
--4.0 & 0.794 &  0.942 & 1.115 &  0.791 & 0.215 &     & 0.133 &  0.180 & 0.275 &  0.370 & 0.207  \\
--4.5 & 0.462 &  0.602 & 0.815 &  0.687 & 0.196 &     & 0.009 &  0.014 & 0.023 &  0.035 & 0.022  \\
\sidehead{}
 & \multicolumn{5}{c}{[\ion{O}{1}]/H$\alpha$}   &     & \multicolumn{5}{c}{[\ion{S}{2}]/H$\alpha$}   \\
\noalign{\vskip -8pt}
 & \multicolumn{5}{c}{\hrulefill}               &     & \multicolumn{5}{c}{\hrulefill}   \\
--3.0 & 0.121 &  0.140 & 0.193 &  0.307 & 0.484 &     & 0.295 &  0.287 & 0.190 &  0.087 & 0.045  \\
--3.5 & 0.170 &  0.200 & 0.269 &  0.400 & 0.556 &     & 0.595 &  0.587 & 0.381 &  0.136 & 0.046  \\
--4.0 & 0.206 &  0.254 & 0.363 &  0.545 & 0.694 &     & 0.860 &  0.898 & 0.645 &  0.230 & 0.061  \\
--4.5 & 0.167 &  0.231 & 0.388 &  0.676 & 0.893 &     & 0.826 &  0.963 & 0.843 &  0.353 & 0.090  \\
\enddata
\tablenotetext{a}{The input SED for the models of this table has $\alpha_{\rm ox}=1.3$. 
See \S4 of the text for details.}
\end{deluxetable}

\clearpage

\clearpage

\clearpage

\clearpage

\end{document}